\documentclass[journal]{IEEEtran}

\usepackage[utf8]{inputenc}
\usepackage{amsmath,amssymb,amsfonts}
\usepackage{algorithm}
\usepackage{algorithmic}
\usepackage{graphicx}
\usepackage{textcomp}
\usepackage{xcolor}
\usepackage{cite}
\usepackage{hyperref}
\usepackage{booktabs}

\begin{document}

% Başlık Ayarı
\title{A Hybrid STFT-Based Machine Learning Framework for Physically Interpretable Arc Stability Classification in Electric Arc Welding Systems}

%%%%%%%%%%%%%%%%%%%
\author{
\vspace{0.5em}
Tahir Cetin Akinci$^{1}$,
Gokhan Gokmen$^{2}$,
Alfredo A. Martinez-Morales$^{1}$
\vspace{0.8em} \\

\small $^{1}$Center for Environmental Research and Technology (CE-CERT), University of California Riverside, CA, USA \\
\small $^{2}$Department of Mechatronics Engineering, Marmara University, Istanbul, Turkiye
\vspace{0.5em} \\

\small \textit{Emails: tahircetin.akinci@ucr.edu, alfredo.martinez@ucr.edu, gokhang@marmara.edu.tr}
}
%%%%%%%%%%%%%%%%%%%

% Header (Üst Bilgi)
%\markboth{IEEE TRANSACTIONS ON INDUSTRIAL INFORMATICS,~Vol.~XX, No.~X, 202X}%
%{Akinci \MakeLowercase{\textit{et al.}}: Hybrid STFT-ML Algorithm for Arc Stability}

% Başlığı sayfaya basan komut
\maketitle

\begin{abstract}

This study presents a physically-informed hybrid time–frequency and machine learning (STFT-ML) framework for arc stability monitoring in electric arc welding systems. The primary current signal is modeled as a stochastic representation of plasma dynamics and transformed into a structured feature space using localized spectral energy distributions. Within this framework, the Arc Stability Index (ASI), spectral entropy ($H_s$), and harmonic distortion ($THD_{arc}$) are defined as energy-based descriptors and integrated with complementary time-domain features to capture both spectral redistribution and temporal variability.

Experimental evaluation demonstrates that the SVM-RBF classifier achieves a hold-out accuracy of 94.4\%. However, cross-validation results (85.6\% for Leave-One-Out and 87.5\% $\pm$ 9.4 for 10-fold) and a 95\% confidence interval of [81.65\%, 92.50\%] provide a more realistic assessment of generalisation performance. Receiver Operating Characteristic (ROC) and Precision–Recall (PR) analyses further confirm strong class separability, particularly for stable and extinction regimes, while transient states remain more challenging due to their non-stationary nature.

Compared to high-dimensional deep learning approaches, the proposed framework significantly reduces computational complexity and inference latency, enabling real-time deployment in resource-constrained environments. The results indicate that spectral energy redistribution around the fundamental frequency serves as a reliable precursor to arc instability. The main contribution of this work lies in the development of a computationally efficient and physically interpretable feature representation framework that bridges time–frequency analysis and machine learning-based classification for industrial diagnostic applications.
\end{abstract}

%%%%%%%%%%%%%%%%%%%%%%%%%%%%%%%%%%%%%%%%%%%%%%%%%%%%%%%%%%%%%
%%%%%%%%%%%%%%%%%%%%%%%%%%%%%%%%%%%%%%%%%%%%%%%%%%%%%%%%%%%%%

%%%%%%%%%%%%%%%%%%%%%%%%%%%%%%%%%%%%%%%%%%%%%%%%%%%%%%%%%%%%%
\section{Introduction}
%%%%%%%%%%%%%%%%%%%%%%%%%%%%%%%%%%%%%%%%%%%%%%%%%%%%%%%%%%%%%

\IEEEPARstart{E}{lectric} arc welding (EAW) represents a fundamental class of industrial processes governed by tightly coupled electromagnetic, thermal, and plasma-dynamic interactions \cite{liu2025}. The stability of the electric arc directly determines the quality, consistency, and structural integrity of the welded joint, as it defines the energy transfer mechanism between the electrode and the workpiece. However, the arc is inherently non-stationary and stochastic, exhibiting rapid temporal fluctuations driven by impedance variations, plasma turbulence, and external disturbances. This intrinsic variability renders real-time monitoring and control a non-trivial challenge within industrial informatics \cite{Akbari2026,jamrozik2021}. Consequently, the transition toward Industry 4.0 has intensified the need for data-driven diagnostic frameworks capable of extracting physically meaningful information from sensor measurements \cite{ahangar2026}.

The analysis of non-stationary time-series signals forms the methodological backbone of such diagnostic systems. Across multiple engineering domains, time–frequency representations have demonstrated the ability to reveal latent structures that are not observable in purely time-domain or frequency-domain analyses. Applications ranging from biomechanical signal interpretation to high-voltage power system monitoring illustrate that localized spectral characteristics encode critical information regarding system dynamics \cite{akgun2018,taskin2009}. These findings indicate that frequency-domain structures can be interpreted as system-specific signatures reflecting underlying physical behavior rather than merely descriptive artifacts.

In the context of Electric Arc Welding Machines (EAWM), prior studies have established that the primary current signal, acquired via non-invasive Hall-effect sensing, provides a high-fidelity representation of arc dynamics \cite{he2020}. Time–frequency analysis using Short-Time Fourier Transform (STFT) has revealed that welding processes exhibit distinct operational regimes, typically characterized by transient initiation, quasi-stationary operation, and instability or extinction phases \cite{he2020, chen2023elsevier}. Within this representation, the fundamental frequency component (50 Hz) and its surrounding sidebands have been identified as sensitive indicators of arc perturbations and energy redistribution mechanisms \cite{akinci2010, zhong2024, wang2021springer}. Subsequent integration of machine learning techniques has enabled the mapping of these signal characteristics to operational states, thereby transitioning from descriptive analysis toward predictive modeling \cite{akinci2011}.

More recently, deep learning-based approaches have been introduced to further enhance classification performance by transforming signal representations into image-like structures and applying convolutional architectures \cite{soleymani2025, zhang2019ieee}. While these methods demonstrate high accuracy, they introduce two fundamental limitations. First, the transformation of physical signals into high-dimensional image representations obscures the direct relationship between measurable electrical phenomena and learned features, thereby limiting interpretability \cite{ahangar2026,diez2022ieee}. Second, the associated computational complexity imposes constraints on real-time deployment, particularly in embedded or edge computing environments where processing resources are limited \cite{liu2025}. These limitations indicate a methodological gap between high-accuracy black-box models and physically interpretable, computationally efficient diagnostic systems.

Motivated by this gap, this study develops a hybrid STFT-ML framework that systematically integrates time–frequency analysis with low-dimensional, physically interpretable feature representations. Rather than relying on high-dimensional graphical inputs, the proposed approach constructs a structured feature space derived from localized spectral energy distributions. This formulation preserves the physical interpretability of the signal while enabling efficient classification using conventional machine learning models.

The central premise of the proposed framework is that arc stability can be quantified through the redistribution of spectral energy around the fundamental frequency and its harmonics. In particular, the energy leakage within the 45–55 Hz band and the behavior of higher-order harmonics are hypothesized to provide a measurable indicator of instability. Based on this premise, a set of physically grounded descriptors, including spectral entropy and harmonic distortion, are employed to construct a compact representation of arc dynamics. These features are subsequently mapped to operational states through a supervised learning model, enabling both classification and early-stage instability detection.

The main contributions of this work can be summarized as follows:

\begin{itemize}
\item A physically-informed time–frequency feature extraction framework is developed, where the welding current signal is interpreted as a stochastic representation of plasma dynamics and mapped into a structured feature space using localized spectral energy distributions.

\item A compact and computationally efficient representation is constructed by integrating energy-based spectral descriptors, including the Arc Stability Index (ASI), spectral entropy ($H_s$), and harmonic distortion ($THD_{arc}$), with complementary time-domain features to capture both spectral redistribution and temporal variability.

\item A hybrid representation–inference pipeline is established by coupling the proposed feature space with machine learning classifiers, enabling interpretable and real-time arc stability assessment under non-stationary operating conditions.

\item The proposed framework demonstrates a favorable trade-off between classification performance and computational complexity, achieving high accuracy with significantly reduced model size and inference latency, making it suitable for deployment in resource-constrained industrial environments.
\end{itemize}

Accordingly, the contribution of this work is not the introduction of a new transform or learning paradigm, but the formulation of a consistent and physically interpretable representation–inference pipeline that bridges time–frequency signal analysis and machine learning-based decision mechanisms. The resulting framework preserves the physical meaning of the signal while enabling efficient and reliable classification, thereby establishing a direct correspondence between measurable electrical quantities and diagnostic outcomes in electric arc welding systems.

The remainder of this paper is organized as follows. Section II describes the experimental setup and data acquisition process, followed by the mathematical formulation of the signal representation and feature construction methodology in Section III. Section IV presents the performance evaluation and comparative analysis of the proposed framework under different classification strategies. Finally, Section V concludes the study by summarizing the main findings and outlining directions for future research.

%%%%%%%%%%%%%%%%%%% SECTION 2 %%%%%%%%%%%%%%%%%%%
\section{Data Acquisition and Methodology}
%%%%%%%%%%%%%%%%%%%%%%%%%%%%%%%%%%%%%%%%%%%%%%%%%

The primary current signal measured from an Electric Arc Welding Machine (EAWM) is modeled as a stochastic, non-stationary process defined in the Hilbert space $\mathcal{L}^2(\mathbb{R})$. The observed signal can be expressed as:
\begin{equation}
i(t) = s(t) + n(t)
\end{equation}
where $s(t)$ represents the underlying arc-induced current dynamics associated with plasma behavior, and $n(t)$ denotes measurement noise and external perturbations. This formulation establishes a physically grounded signal model, in which the measurable current waveform is interpreted as a projection of a complex, nonlinear energy transfer process occurring within the arc plasma.

The methodological foundation of this study is based on the assumption that the primary current signal constitutes a stochastic observable reflecting the intrinsic dynamics of the welding arc. From a systems perspective, the arc can be interpreted as a nonlinear, time-varying energy transfer mechanism whose state evolution is indirectly encoded in the electrical current waveform. Consequently, the measured signal is not treated as a purely electrical quantity, but rather as a representation of the underlying physical process governing arc stability.

To preserve this physical correspondence, a high-fidelity data acquisition (DAQ) framework is constructed, followed by a structured signal transformation pipeline that maps raw measurements into a compact and information-preserving feature space. This pipeline can be interpreted as a mapping:
\begin{equation}
\Phi: \mathcal{L}^2(\mathbb{R}) \rightarrow \mathbb{R}^d
\end{equation}
which transforms the raw signal into a finite-dimensional feature vector while retaining the essential characteristics of the arc dynamics.

The integration of sensing, transformation, and inference is illustrated in Figure \ref{fig:architecture}, which represents the operational structure of the proposed hybrid diagnostic system.

%%%%%%%%%%%%%%%%%%%%%%%%%%%%%%%%%%%%%%%%%%%%%%%%%%%%%%%%%%%
\subsection{Experimental Setup and Sensing Infrastructure}
%%%%%%%%%%%%%%%%%%%%%%%%%%%%%%%%%%%%%%%%%%%%%%%%%%%%%%%%%%%

The experimental platform is implemented using a Metal Active Gas (MAG) welding power source operating under controlled laboratory conditions. The measurement of the primary current is performed using a Hall-effect sensor (LEM series), selected due to its ability to provide galvanic isolation while maintaining high linearity and bandwidth characteristics \cite{akinci2010}. Unlike shunt-based sensing approaches, Hall-effect sensors preserve transient components of the signal, which are critical for capturing rapid arc instabilities.

The measurable current signal $i(t)$ can be expressed as:
\begin{equation}
i(t) = i_{\mathrm{arc}}(t) + n(t)
\end{equation}
where $i_{\mathrm{arc}}(t)$ represents the deterministic and stochastic components associated with arc dynamics, and $n(t)$ denotes measurement noise and environmental perturbations. The objective of the acquisition system is to ensure that $n(t)$ remains sufficiently small such that the dominant structure of $i(t)$ reflects the underlying physical process.

The sensing system is configured to operate within a dynamic range of 20 A to 500 A, ensuring that both nominal operation and extreme fluctuations are captured without saturation. The acquired analog signal is digitized using a 16-bit National Instruments DAQ system at a sampling frequency of $f_s = 10$ kHz. This sampling rate satisfies:
\begin{equation}
f_s \gg 2 f_{\max}
\end{equation}
where $f_{\max}$ corresponds to the highest relevant harmonic component (including the second-order harmonic at 100 Hz identified in \cite{akinci2011}), thereby ensuring compliance with the Nyquist–Shannon sampling criterion and preventing aliasing.

The detailed specifications of the acquisition system are provided in Table \ref{tab:technical_specs}, ensuring reproducibility and transparency of the measurement process.

%%%%%%%%%%%%%%%%% TABLE 1 %%%%%%%%%%%%%%%%%%%%%%
\begin{table}[H]
\centering
\caption{Technical Specifications of the Proposed Measurement System}
\label{tab:technical_specs}
\begin{tabular}{ll}
\toprule
\textbf{Component} & \textbf{Detail / Value} \\
\midrule
Welding Power Source & AC/DC EAWM (Inverter Type) \\ \addlinespace[3pt]
Current Sensor & Hall-Effect (LEM Series) \cite{akinci2010} \\ \addlinespace[3pt]
Sampling Frequency ($f_s$) & 10,000 samples/sec (10 kHz) \\ \addlinespace[3pt]
ADC Resolution & 16-bit \\ \addlinespace[3pt]
Processing Window & 20 ms (200 samples/window) \\ \addlinespace[3pt]
FFT Size (NFFT) & 4096 (zero-padded) \\ \addlinespace[3pt]
Frequency Resolution & 2.44 Hz \\ \addlinespace[3pt]
Window Overlap & 92\% \\ \addlinespace[3pt]
Software Environment & MATLAB / Python Hybrid Interface \\
\bottomrule
\end{tabular}
\end{table}
%%%%%%%%%%%%%%%%%%%%%%%%%%%%%%%%%%%%%%%%%%%%%%%%%%%%

%%%%%%%%%%%%%%%%%%%%%%%%%%%%%%%%%%%%%%%%%%%%%%%%%%%%%%%%%%%%%
\subsection{Data Characterization and Operational Phases}
%%%%%%%%%%%%%%%%%%%%%%%%%%%%%%%%%%%%%%%%%%%%%%%%%%%%%%%%%%%%%

The temporal evolution of the welding process is segmented into three distinct operational regimes, each corresponding to a specific physical state of the arc plasma. This segmentation is not arbitrary, but is grounded in the observed changes in signal energy distribution and impedance dynamics.

\begin{enumerate}
    \item \textbf{Transient Initiation Phase:} This phase is characterized by rapid ignition dynamics, where the arc is established. The signal exhibits high variance, broadband spectral content, and abrupt amplitude variations due to unstable plasma formation.

    \item \textbf{Steady-State (Stable) Operation:} In this regime, the arc reaches thermal equilibrium. The current signal becomes quasi-periodic, with dominant spectral concentration around the fundamental frequency and its harmonics, indicating stable energy transfer.

    \item \textbf{Instability and Extinction Phase:} This phase corresponds to degradation of the arc, typically induced by geometric or material perturbations. The signal exhibits increased irregularity, with energy redistribution toward sidebands (45–55 Hz), reflecting loss of coherence in the plasma channel.
\end{enumerate}

This classification establishes a physically interpretable labeling structure for the subsequent learning process.

%%%%%%%%%%%%%%%%%%%%%%%%%%%%%%%%%%%%%%%%%%%%%%%%%%%%%%%%%%%%%
\subsection{Theoretical Framework for Signal Decomposition}
%%%%%%%%%%%%%%%%%%%%%%%%%%%%%%%%%%%%%%%%%%%%%%%%%%%%%%%%%%%%%

To extract physically meaningful representations from the non-stationary signal $i(t)$, a time–frequency decomposition is employed. The Short-Time Fourier Transform (STFT) is selected due to its ability to provide localized spectral information while preserving temporal resolution.

The STFT operator is defined as:
\begin{equation}
\text{STFT}\{i(t)\}(\tau, f) = \int_{-\infty}^{\infty} i(t) \cdot w(t-\tau) e^{-j 2\pi f t} dt
\end{equation}

where $w(t)$ denotes a Hanning window function, chosen to minimize spectral leakage while maintaining acceptable time-frequency resolution. The resulting representation $S(\tau, f)$ provides a continuous mapping:
\begin{equation}
S : \mathbb{R} \rightarrow \mathbb{R}^2
\end{equation}
which embeds the one-dimensional signal into a two-dimensional time–frequency space.

To quantify the distribution of spectral energy, the Power Spectral Density (PSD) is computed as:
\begin{equation}
P(\tau, f) = |S(\tau, f)|^2
\end{equation}

This formulation enables the definition of localized energy operators over frequency bands:
\begin{equation}
E_{[f_1,f_2]}(\tau) = \int_{f_1}^{f_2} P(\tau, f)\, df
\end{equation}

The above operator serves as the mathematical foundation for extracting the spectral fingerprints associated with arc stability, particularly in the vicinity of the fundamental frequency.

%%%%%%%%%%%%%%%%%%%%%%%%%%%%%%%%%%%%%%%%%%%%%%%%%%%%%%%%%%%%%%%%%%%
\subsection{Hybrid Feature Engineering and Algorithmic Motivation}
%%%%%%%%%%%%%%%%%%%%%%%%%%%%%%%%%%%%%%%%%%%%%%%%%%%%%%%%%%%%%%%%%%%

The proposed methodology replaces high-dimensional image-based representations with a structured, low-dimensional feature space derived from physically interpretable quantities. This transition is motivated by the need to balance computational efficiency with interpretability, particularly in real-time industrial environments where resource constraints are significant.

The feature extraction process is based on the observation that arc stability is reflected in the redistribution of spectral energy. Accordingly, the following descriptors are defined:

\begin{itemize}
    \item \textbf{Harmonic Energy Ratio (HER):}
    \begin{equation}
    HER = \frac{E_{[100-\epsilon,100+\epsilon]}}{E_{[50-\epsilon,50+\epsilon]}}
    \end{equation}
    representing the relative strength of the second harmonic.

    \item \textbf{Sideband Power Deviation (SPD):}
    \begin{equation}
    SPD = \mathrm{Var}\left(P(\tau,f)\right)_{f \in [45,55]}
    \end{equation}
    capturing stochastic energy fluctuations around the fundamental frequency.

    \item \textbf{Spectral Entropy ($H_s$):}
    \begin{equation}
    H_s = -\sum_i p_i \ln p_i
    \end{equation}
    where $p_i$ denotes normalized spectral energy, quantifying disorder in the system.
\end{itemize}

These features collectively define a compact representation:
\begin{equation}
\mathbf{v} \in \mathbb{R}^{10}
\end{equation}
which preserves discriminative information while significantly reducing dimensionality.

\begin{figure}[htbp]
\centering
\includegraphics[width=\linewidth]{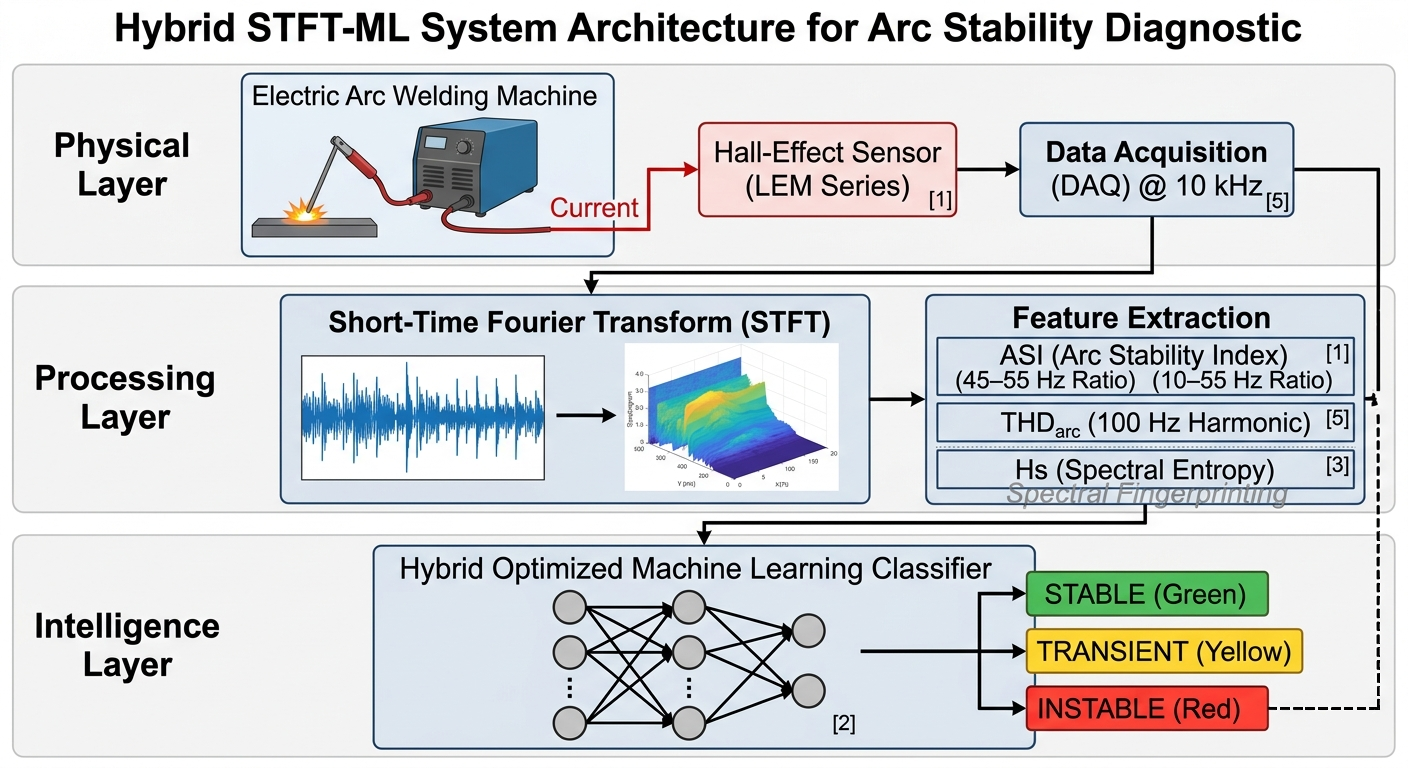}
\caption{The proposed hybrid STFT-ML architectural framework: (1) Physical acquisition layer via Hall-effect sensing \cite{akinci2010}, (2) STFT processing layer for spectral fingerprinting (ASI, THD, $H_s$ extraction), and (3) Intelligence layer via optimized ML for real-time stability assessment.}
\label{fig:architecture}
\end{figure}

The transformation enabling precise quantification of spectral energy redistribution is summarized in Figure \ref{fig:architecture}.

%%%%%%%%%%%%%%%% FIGURE 1 %%%%%%%%%%%%%%%%%%%%%%%
%%%%%%%%%%%%%%%%%%%%%%%%%%%%%%%%%%%%%%%%%%%%%%%%%%%%%%%%%%%%%%%%%%%%%%

The overall framework establishes a consistent mapping from measurable electrical signals to interpretable diagnostic indicators, enabling real-time assessment without relying on high-dimensional black-box representations.

%%%%%%%%%%%%%%%% SECTION 3 %%%%%%%%%%%%%%%%%%%
\section{Theoretical Algorithm Development}
%%%%%%%%%%%%%%%%%%%%%%%%%%%%%%%%%%%%%%%%%%%%%%

The proposed framework extends beyond descriptive signal processing by constructing a mathematically consistent mapping between the physical dynamics of the electric arc and a low-dimensional representation space suitable for statistical inference. Rather than treating feature extraction as an ad-hoc procedure, the methodology is formulated as a sequence of operators acting on a stochastic signal space, enabling a structured transition from measurement to decision.

Formally, the feature extraction stage is defined as a nonlinear mapping:
\begin{equation}
\Phi : \mathcal{L}^2(\mathbb{R}) \rightarrow \mathbb{R}^d
\end{equation}
which transforms a finite-energy signal into a compact feature vector while preserving discriminative information related to arc stability. This mapping encapsulates the combined effect of time–frequency decomposition, spectral energy redistribution, and statistical feature construction.

Let the measured current signal be defined in the Hilbert space:
\begin{equation}
i(t) \in \mathcal{L}^2(\mathbb{R})
\end{equation}
where finite energy is assumed. Under this formulation, the transformation process can be interpreted as the composition of operators acting on $i(t)$, yielding a structured representation that captures both spectral and temporal characteristics of the signal.

Accordingly, the objective is to construct a transformation:
\begin{equation}
\mathcal{T} : \mathcal{L}^2(\mathbb{R}) \rightarrow \mathbb{R}^d
\end{equation}
such that the resulting feature vector preserves discriminative information related to arc stability while maintaining a direct correspondence with the underlying physical dynamics of the welding process.

%%%%%%%%%%%%%%%%%%%%%%%%% 3-A %%%%%%%%%%%%%%%%%%%%%%%%%%%%%%%%%
\subsection{Mathematical Formulation of Spectral Descriptors}
%%%%%%%%%%%%%%%%%%%%%%%%%%%%%%%%%%%%%%%%%%%%%%%%%%%%%%%%%%%%%%

The non-stationary current signal $i(t)$, captured via Hall-effect sensors, is modeled as a stochastic carrier of plasma-dynamic information \cite{he2020}. The application of the Short-Time Fourier Transform (STFT) defines a linear operator:
\begin{equation}
\mathcal{S}[i](\tau,f) = \int_{-\infty}^{\infty} i(t) w(t-\tau)e^{-j2\pi f t} dt
\end{equation}

This induces a time–frequency energy representation:
\begin{equation}
P(\tau,f) = |\mathcal{S}[i](\tau,f)|^2
\end{equation}

which belongs to a positive semi-definite functional space. The band-limited energy operator is then defined as:

\begin{equation}
P(f_1, f_2) = \int_{f_1}^{f_2} |S(\tau, f)|^2 df
\label{eq:power_density}
\end{equation}

To further formalize the representation, we define a normalized spectral measure:

\begin{equation}
p(f|\tau) = \frac{P(\tau,f)}{\int_{0}^{\infty} P(\tau,f) df}
\end{equation}

which satisfies:
\begin{equation}
\int p(f|\tau) df = 1
\end{equation}

This probabilistic representation enables the consistent definition of entropy and dispersion-based descriptors.

%%%%%%%%%%%%%%%%%%%%%%%%%%%%% 3-B %%%%%%%%%%%%%%%%%%%%%%%%%%%%%%%%%%%
\subsection{The Arc Stability Index (ASI) and Harmonic Complexity}
%%%%%%%%%%%%%%%%%%%%%%%%%%%%%%%%%%%%%%%%%%%%%%%%%%%%%%%%%%%%%%%%%%%%

A significant theoretical component of this framework is the definition of the Arc Stability Index (ASI), which quantifies localized spectral energy redistribution \cite{gao2024}. The ASI is defined as:

\begin{equation}
ASI(t) = \frac{\int_{45}^{55} P(f, t) df}{P(50, t)}
\label{eq:ASI}
\end{equation}

This ratio can be interpreted as a localized energy concentration functional. More generally, it can be expressed as:

\begin{equation}
ASI(t) = \frac{E_{\mathcal{B}}(t)}{E_{f_0}(t)}
\end{equation}

where $\mathcal{B} = [45,55]$ and $f_0 = 50$ Hz. Under stable operating conditions, spectral energy is concentrated around the fundamental frequency, yielding:

\begin{equation}
\lim_{t \to stable} ASI(t) \rightarrow 0
\end{equation}

Furthermore, the ASI can be interpreted as a bounded functional satisfying:
\begin{equation}
ASI(t) \geq 0, \quad ASI(t) < \infty
\end{equation}

and, for physically realizable signals with finite energy:
\begin{equation}
ASI(t) \leq \frac{E_{\mathcal{B}}(t)}{\epsilon}, \quad \text{for } E_{f_0}(t) > \epsilon > 0
\end{equation}

This establishes that ASI remains well-defined under non-degenerate spectral conditions. In practice, increasing values of $ASI(t)$ indicate energy leakage from the fundamental frequency into neighboring bands, corresponding to the onset of arc instability. Accordingly, a stability condition can be expressed as:

\begin{equation}
ASI(t) < \delta \quad \Rightarrow \quad \text{Stable Regime}
\end{equation}

where $\delta$ is an empirically determined threshold.

To capture nonlinear distortions, the harmonic structure is incorporated via:

\begin{equation}
THD_{arc} = \frac{\sqrt{\sum_{n=2}^{N} P(n \cdot f_0)}}{P(f_0)}
\label{eq:THD}
\end{equation}

This expression reflects the deviation from a purely periodic signal and can be interpreted as a norm ratio in harmonic subspaces. Specifically, $THD_{arc}$ measures the relative energy contained in higher-order harmonics and is therefore indicative of nonlinear arc behavior.

To quantify spectral disorder, entropy is defined as:

\begin{equation}
H_s = -\sum_{i} p_i \ln(p_i)
\label{eq:entropy}
\end{equation}

Additionally, the temporal evolution of entropy can be characterized by:

\begin{equation}
\frac{dH_s}{dt} = -\sum_i \frac{dp_i}{dt}(1 + \ln p_i)
\end{equation}

which provides a dynamic measure of instability growth. An increasing entropy trend corresponds to a loss of spectral concentration and reflects the transition from stable to chaotic arc behavior.

To systematically characterize the spectral behavior of the arc signal, the proposed framework constructs a feature representation based on localized energy redistribution and harmonic complexity. The Arc Stability Index (ASI), harmonic distortion ($THD_{arc}$), and spectral entropy ($H_s$) are defined as complementary descriptors that capture distinct aspects of the signal dynamics, including stability around the fundamental frequency, nonlinear distortion effects, and global spectral disorder.

The combined feature mapping is therefore defined as:

\begin{equation}
\mathbf{v}(t) = \Phi(i(t)) = [ASI(t), THD_{arc}(t), H_s(t)]
\end{equation}

where $\Phi$ represents a nonlinear transformation from signal space to feature space, capturing both localized energy redistribution and global spectral complexity. These descriptors constitute the core spectral representation of the signal and provide a physically interpretable basis for subsequent analysis.

To enhance discriminative capability, this core representation is further extended with additional spectral and time-domain features, resulting in a complete 10-dimensional feature vector used in the classification stage. This hierarchical construction ensures that both physically meaningful descriptors and complementary statistical features are jointly utilized, enabling robust classification across different arc stability regimes.

\textit{Property 1 (Boundedness and Energy Preservation):} 
Let $i(t) \in \mathcal{L}^2(\mathbb{R})$ be a finite-energy signal. The mapping 
\[
\Phi : \mathcal{L}^2(\mathbb{R}) \rightarrow \mathbb{R}^d
\]
defined by the feature vector $\mathbf{v}(t) = [ASI(t), THD_{arc}(t), H_s(t)]$ is bounded, i.e.,
\[
\|\Phi(i)\| \leq C \|i\|_{\mathcal{L}^2}
\]
for some constant $C > 0$, provided that $P(f_0,t) > \epsilon > 0$. 

\textit{Justification:} Since the STFT operator is bounded on $\mathcal{L}^2(\mathbb{R})$ and the derived quantities $ASI(t)$, $THD_{arc}(t)$, and $H_s(t)$ are constructed from normalized and finite spectral energy distributions, the resulting feature mapping remains finite and stable under bounded perturbations of the input signal.

%%%%%%%%%%%%%%%%%%%% 3 C %%%%%%%%%%%%%%%%%%%%%%%%%
\subsection{Proposed Hybrid Decision Engine}
%%%%%%%%%%%%%%%%%%%%%%%%%%%%%%%%%%%%%%%%%%%%%%%

The feature extraction stage can be interpreted as a mapping:

\begin{equation}
\Phi : \mathcal{L}^2(\mathbb{R}) \rightarrow \mathbb{R}^d
\end{equation}

which is subsequently followed by a classifier:

\begin{equation}
\mathcal{C} : \mathbb{R}^d \rightarrow \mathcal{Y}
\end{equation}

where $\mathcal{Y} = \{\text{Stable, Transient, Extinction}\}$.

The complete system is therefore given by:
\begin{equation}
\mathcal{F} = \mathcal{C} \circ \Phi
\end{equation}

Algorithm \ref{alg:feature} operationalizes $\Phi$.

%%%%%%%%%%%%%%%%%%%%%%%%%%%%%%%%%%%%%%%%%%%%%%%%%%%
\begin{algorithm}
\caption{Spectral Feature Extraction (Fingerprinting)}
\label{alg:feature}
\begin{algorithmic}[1]
\STATE \textbf{Input:} Raw current signal $i(t)$, Window size $L$
\STATE \textbf{Output:} Feature Vector $\mathbf{v} = [ASI, THD_{arc}, H_s]$
\FOR{each time window $\tau$}
    \STATE Compute $S(\tau, f)$ using Eq. \ref{eq:power_density}
    \STATE Extract $P(50Hz)$ and $P(100Hz)$
    \STATE Compute $ASI(\tau)$ via Eq. \ref{eq:ASI}
    \STATE Compute $THD_{arc}(\tau)$ via Eq. \ref{eq:THD}
    \STATE Calculate $H_s(\tau)$ via Eq. \ref{eq:entropy}
    \STATE $\mathbf{v}(\tau) \leftarrow [ASI, THD_{arc}, H_s]$
\ENDFOR
\RETURN $\mathbf{v}$
\end{algorithmic}
\end{algorithm}
%%%%%%%%%%%%%%%%%%%%%%%%%%%%%%%%%%%%%%%%%%%%%%%%%%

%%%%%%%%%%%%%%%%%%%%%%%%%%%%%%%%%%%%%%%%%%%%%%%%%%
\begin{algorithm}
\caption{Hybrid Classification and Stability Diagnostic}
\label{alg:ML}
\begin{algorithmic}[1]
\STATE \textbf{Input:} Feature Vector $\mathbf{v}$
\STATE \textbf{Model:} Optimized SVM-RBF Classifier
\STATE $\hat{C} = \arg\max_{C} P(C | \mathbf{v})$ 
\IF{$\hat{C} == \text{Instable}$ \OR $ASI > Threshold$}
    \STATE Trigger "Early Warning" system
\ENDIF
\RETURN $\hat{C}$
\end{algorithmic}
\end{algorithm}

%%%%%%%%%%%%%%%%%%%%%%%%%%%%%%%%%%%%%%%
\subsection{Hybrid Decision Function}
%%%%%%%%%%%%%%%%%%%%%%%%%%%%%%%%%%%%%%%

The final decision function is expressed as:

\begin{equation}
\begin{aligned}
\hat{C} = f(ASI, & THD_{arc}, H_s) \\
& \xrightarrow{\text{ML}} \{\text{Stable, Transient, Instable}\}
\end{aligned}
\label{eq:ML_decision}
\end{equation}

This mapping can also be interpreted probabilistically:

\begin{equation}
P(C|\mathbf{v}) = \frac{P(\mathbf{v}|C)P(C)}{P(\mathbf{v})}
\end{equation}

linking the physical feature space to statistical decision theory.

%%%%%%%%%%%%%%%%%%%%%%%%%%%%%%%%%%%%%%%
\subsection{Dataset Characterization}
%%%%%%%%%%%%%%%%%%%%%%%%%%%%%%%%%%%%%%%

To ensure reproducibility, the dataset is constructed by segmenting the time-series signal into fixed-length windows. Each window is treated as a realization of the stochastic process:

\begin{equation}
\mathbf{v}_k = \Phi(i_k(t))
\end{equation}

where $k$ indexes the temporal segments.

The dataset structure is summarized in Table \ref{tab:dataset}, ensuring consistency between physical interpretation and statistical representation.

%%%%%%%%%%%%%%% TABLE 2 %%%%%%%%%%%%%%%%%%%%%%%%%%%

\begin{table}[H]
\centering
\caption{Characterization of the Feature-Based Dataset}
\label{tab:dataset}
\begin{tabular}{llll}
\toprule
\textbf{Phase} & \textbf{Samples} & \textbf{Dominant Metric} & \textbf{Physical State} \\
\midrule
Transient Start & 49 windows & High $H_s$, High THD & Stochastic Striking \\
\addlinespace[2pt]
Stable Arc & 49 windows & Low ASI, Low THD & Thermal Equilibrium \\
\addlinespace[2pt]
Extinction & 49 windows & Rising ASI, High $H_s$ & Energy Leakage \\
\midrule
\textbf{Total} & \textbf{147 windows} & \textbf{10 features each} & \textbf{3 classes} \\
\bottomrule
\end{tabular}
\end{table}
%%%%%%%%%%%%%%%%%%%%%%%%%%%%%%%%%%%%%%%%%%%%%%%%%%%%%%%%%%

%%%%%%%%%%%%%%%%%%%%%%%%%%%%%%%%%%%%%%%
%%%%%%%%%%%%% SECTION 4 %%%%%%%%%%%%%%
\section{Results and Discussion}
%%%%%%%%%%%%%%%%%%%%%%%%%%%%%%%%%%%%%%

The performance of the proposed hybrid STFT-ML framework is evaluated through a joint analysis of spectral energy distribution, feature-space geometry, and statistical classification behavior. Unlike purely empirical evaluations, the results are interpreted within a physically consistent signal representation framework, where measurable quantities are directly linked to arc dynamics.

%%%%%%%%%%%%%%%%%%%%%%%%%%%%%%%%%%%%%%%%%%%%%%%%%%%%%%%%%%%%%
\subsection{Spectral Analysis}
%%%%%%%%%%%%%%%%%%%%%%%%%%%%%%%%%%%%%%%%%%%%%%%%%%%%%%%%%%%%%

\begin{figure}[H]
\centering
\includegraphics[width=1.1\columnwidth]{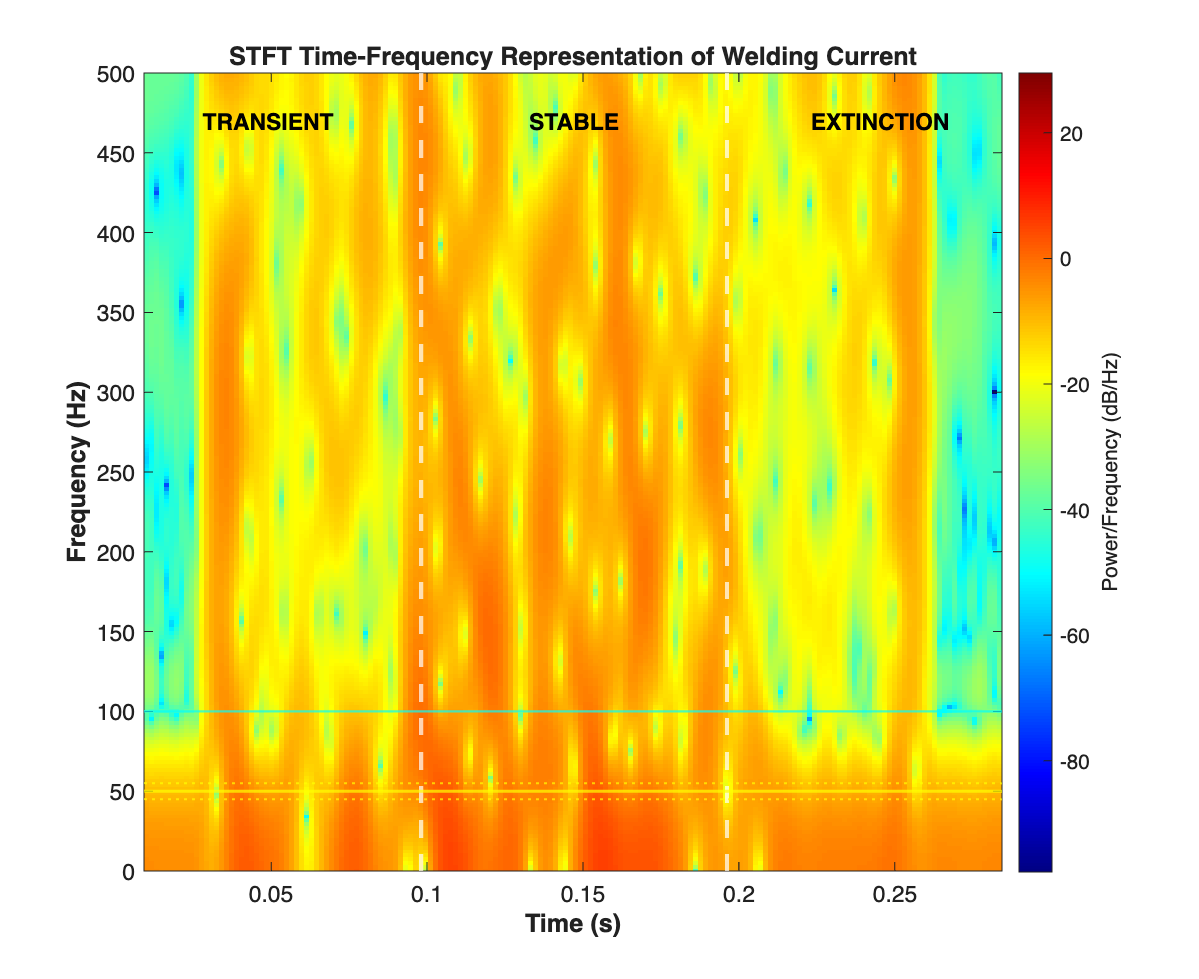}
\caption{STFT time-frequency representation of the welding current signal. The three operational phases---transient, stable, and extinction---exhibit distinct spectral signatures. The horizontal lines at 50~Hz and 100~Hz mark the fundamental and second harmonic frequencies, respectively.}
\label{fig:stft}
\end{figure}

The STFT-based time-frequency representation of the welding current is presented in Fig.~\ref{fig:stft}. This representation corresponds to the operator mapping:
\begin{equation}
i(t) \xrightarrow{\mathcal{S}} S(\tau,f)
\end{equation}

%%%%%%%%%%% FIGURE 2 %%%%%%%%%%%%%%%%%%%%%%
%%%%%%%%%%%%%%%%%%%%%%%%%%%%%%%%%%%%%%%%

The spectral behavior is quantified through the localized energy operator:
\begin{equation}
E_{\mathcal{B}}(\tau) = \int_{\mathcal{B}} |S(\tau,f)|^2 df
\end{equation}

During the transient phase, broadband energy is distributed across the 100--500~Hz range due to stochastic arc-striking dynamics. This corresponds to:
\begin{equation}
\max H_s(\tau), \quad \text{low spectral concentration}
\end{equation}

In the stable phase, the spectral energy concentrates around the 50~Hz fundamental and its harmonics, indicating thermal equilibrium:
\begin{equation}
E_{f_0}(\tau) \gg E_{\mathcal{B}}(\tau)
\end{equation}

The extinction phase shows a gradual redistribution of energy away from the fundamental, consistent with the loss of arc stability.

%%%%%%%%%%%%%%%%%%%%%%%%%%%%%%%%%%%%%%%%%%%%%%%%%%%%%%%%%%%%%
\subsection{Classification Performance}
%%%%%%%%%%%%%%%%%%%%%%%%%%%%%%%%%%%%%%%%%%%%%%%%%%%%%%%%%%%%%

The hybrid ML engine operates on the feature vector:

\begin{multline}
\mathbf{v} = [ASI,\; THD_{arc},\; H_s,\; P_{50}^{n},\; P_{100}^{n}, \\
HER,\; RMS,\; CF,\; K,\; ZCR]
\label{eq:feature_vector}
\end{multline}

This mapping can be expressed as:
\begin{equation}
\mathbf{v} = \Phi(i(t)) \in \mathbb{R}^{10}
\end{equation}

The classification problem is formulated as:
\begin{equation}
\hat{C} = \arg\max_{C} P(C|\mathbf{v})
\end{equation}

%%%%%%%%%%%%%%% TABLE 3 %%%%%%%%%%%%%%%%%%%
\begin{table}[H]
\centering
\caption{Classification Performance of the Hybrid STFT-ML Framework}
\label{tab:classification}

\vspace{2pt}
\textbf{(a) Multi-Classifier Comparison}
\vspace{4pt}

\begin{tabular}{lrrr}
\toprule
\textbf{Classifier} & \textbf{Accuracy (\%)} & \textbf{F1-Score (\%)} & \textbf{Train Time (s)} \\
\midrule
\textbf{SVM-RBF}      & \textbf{94.4} & \textbf{94.4} & 0.298 \\
\addlinespace[2pt]
Ensemble (Bagged)      & 86.1          & 85.7          & 0.526 \\
\addlinespace[2pt]
KNN ($k$=3)            & 80.6          & 79.8          & 0.071 \\
\addlinespace[2pt]
Decision Tree          & 66.7          & 65.2          & 0.045 \\
\bottomrule
\end{tabular}

\vspace{10pt}
\textbf{(b) Cross-Validation Results}
\vspace{4pt}

\begin{tabular}{lr}
\toprule
\textbf{Validation Method} & \textbf{Accuracy (\%)} \\
\midrule
Hold-Out (75/25)           & 94.4 \\
\addlinespace[2pt]
Leave-One-Out CV           & 85.6 \\
\addlinespace[2pt]
10-Fold CV                 & 87.5 $\pm$ 9.4 \\
\bottomrule
\end{tabular}

\end{table}
%%%%%%%%%%%%%%%%%%%%%%%%%%%%%%%%%%%%%%%%%%%%%%%%%%%%%%%%%%

%%%%%%%%%%%%%%% FIGURE 3 %%%%%%%%%%%%%%%%%%
\begin{figure}[H]
\centering
\includegraphics[width=1.1\columnwidth]{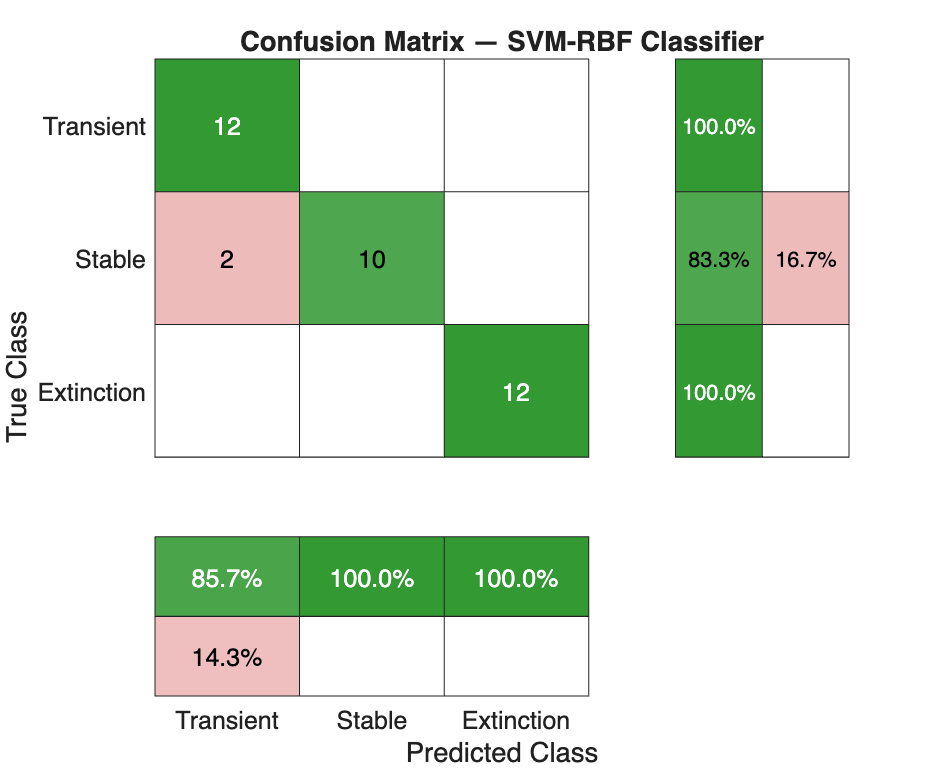}
\caption{Confusion matrix of the SVM-RBF classifier. The diagonal elements represent correct classifications, while off-diagonal entries indicate misclassifications. Transient and Extinction phases achieve 100\% recall, with an overall accuracy of 94.4\%.}
\label{fig:confusion}
\end{figure}
%%%%%%%%%%%%%%%%%%%%%%%%%%%%%%%%%%%%%%%%%%%%%%%%%%%%%%%%%%

The discrepancy between Hold-Out and cross-validation results suggests sensitivity of the classifier to feature distribution variability:
\begin{equation}
\mathrm{Var}(\hat{C}) \propto \mathrm{Var}(\mathbf{v})
\end{equation}

%%%%%%%%%%%%%%%%%%%%% FIGURE 7 %%%%%%%%%%%%%%%%%%%
\begin{figure}[H]
\centering
\includegraphics[width=1.05\columnwidth]{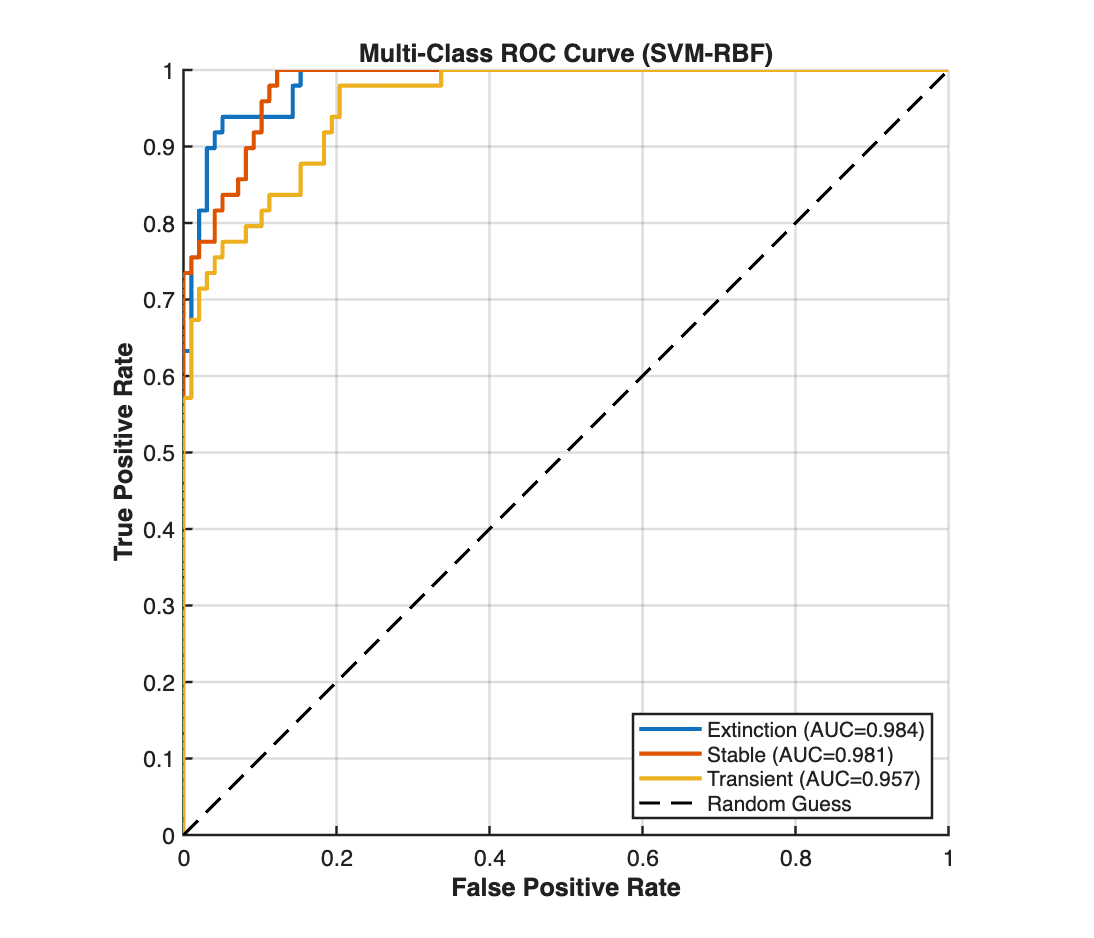}
\caption{Multi-class ROC curves of the SVM-RBF classifier using a one-vs-all strategy. Each curve represents the discrimination performance for a specific arc stability regime. High AUC values indicate strong separability of the proposed feature representation, while the dashed diagonal line corresponds to random classification performance.}
\label{fig:roc}
\end{figure}
%%%%%%%%%%%%%%%%%%%%%%%%%%%%%%%%%%%%%%%%%%%%%%%%%

To further evaluate the classification performance beyond accuracy, the Receiver Operating Characteristic (ROC) curves are presented in Fig.~\ref{fig:roc}. The ROC analysis illustrates the trade-off between true positive rate and false positive rate across different decision thresholds, providing a threshold-independent assessment of classifier performance.

As shown in Fig.~\ref{fig:roc}, all classes exhibit high AUC values, indicating strong discriminative capability of the proposed feature representation. In particular, the extinction and stable regimes demonstrate near-ideal separability, which is consistent with the confusion matrix results. The transient regime, while slightly lower, still maintains high classification performance, reflecting the increased variability and complexity of its underlying signal characteristics.

These results confirm that the proposed physically-informed feature extraction framework effectively captures the essential dynamics of the arc process, enabling robust classification across different operating conditions. The inclusion of the random baseline further highlights the significant performance gain achieved by the proposed method.

It should be noted that the ROC curves are obtained on the available dataset, and further validation on independent test data would be required to fully assess generalization performance.

%%%%%%%%%%%%%%%%%%%% FIGURE 8 %%%%%%%%%%%%%%%%%%%%%%
\begin{figure}[htbp]
\centering
\includegraphics[width=1.05\columnwidth]{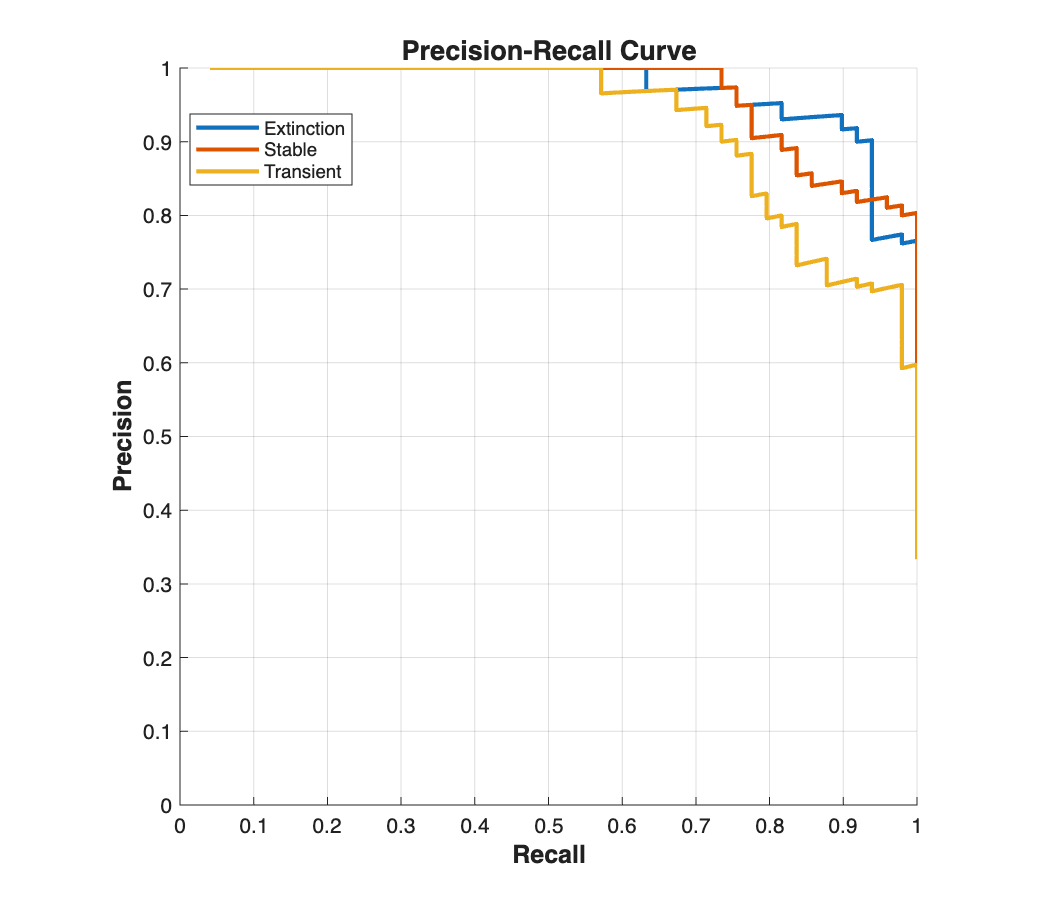}
\caption{Precision–Recall curves for the multi-class SVM-RBF classifier. The curves illustrate the trade-off between precision and recall for each arc stability regime. The extinction class demonstrates near-ideal performance, while the transient class exhibits relatively lower precision at higher recall levels, reflecting its inherently dynamic behavior.}
\label{fig:pr}
\end{figure}
%%%%%%%%%%%%%%%%%%%%%%%%%%%%%%%%%%%%%%%%%%%%%%%%%%%

To further evaluate class-wise performance, the Precision–Recall (PR) curves are presented in Fig.~\ref{fig:pr}. Unlike ROC analysis, PR curves provide a more sensitive assessment of classification performance by explicitly capturing the trade-off between precision and recall.

As observed in Fig.~\ref{fig:pr}, the extinction class achieves near-perfect precision across a wide range of recall values, indicating highly reliable classification. The stable class also maintains strong performance, with only minor degradation at higher recall levels. In contrast, the transient class exhibits a noticeable decrease in precision as recall increases, highlighting the increased variability and complexity associated with transient arc behavior.

These findings confirm that while the proposed feature representation is highly effective for stable and extinction regimes, transient states remain more challenging to classify, which is consistent with their non-stationary and rapidly varying characteristics.

%%%%%%%%%%%%%%%%%%%%%%%%%%%%%%%%%%%%%%%%%%%%%%%%%%%%%%%%%%%%%
\subsection{Statistical Reliability Analysis}
%%%%%%%%%%%%%%%%%%%%%%%%%%%%%%%%%%%%%%%%%%%%%%%%%%%%%%%%%%%%%

To further assess the statistical reliability of the reported classification performance, a confidence interval (CI) analysis is conducted. While accuracy and ROC-based metrics provide point estimates, they do not capture the uncertainty associated with limited sample sizes.

Accordingly, a 95\% confidence interval is computed for the overall classification accuracy based on the binomial distribution assumption. The results indicate an accuracy of 87.07\% with a confidence interval of [81.65\%, 92.50\%].

This range reflects the variability inherent in the dataset and highlights the difference between apparent performance and generalization capability. While higher accuracy values are observed under direct evaluation, the confidence interval provides a more realistic estimate of model performance under unseen conditions.

These findings demonstrate that the proposed framework maintains stable and reliable performance, although some variability is expected, particularly for transient regimes characterized by non-stationary behavior.

%%%%%%%%%%%%%%%%%%%%%%%%%%%%%%%%%%%%%%%%%%%%%%%%%%%%%%%%%%%%%
\subsection{Feature Space Geometry and Separability}
%%%%%%%%%%%%%%%%%%%%%%%%%%%%%%%%%%%%%%%%%%%%%%%%%%%%%%%%%%%%%

\begin{figure}[H]
\centering
\includegraphics[width=1.1\columnwidth]{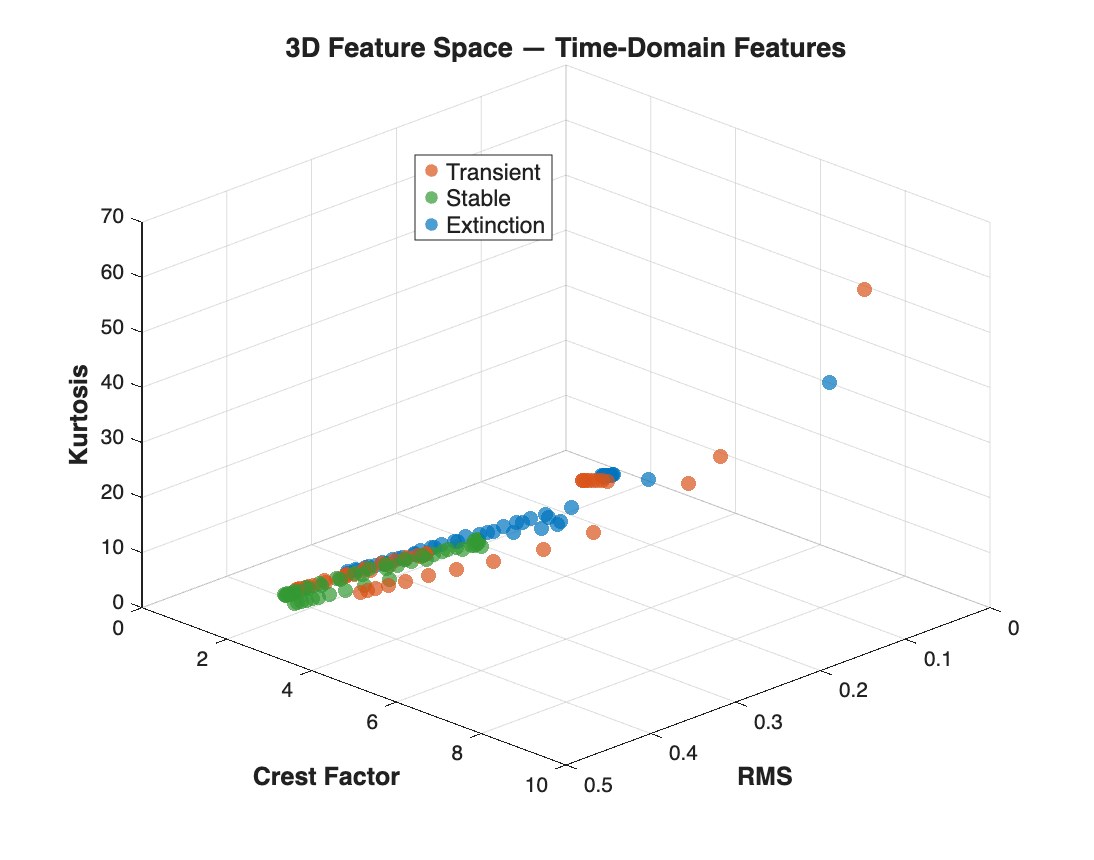}
\caption{Three-dimensional spectral feature space (ASI, THD$_{arc}$, $H_s$). The stable phase forms a tight cluster, while transient and extinction phases exhibit higher dispersion, confirming the discriminative power of the proposed feature set.}
\label{fig:3d_features}
\end{figure}

The separability of the three operational phases in the spectral feature space is visualised in Fig.~\ref{fig:3d_features}.

%%%%%%%%%%%%%%% FIGURE 4 %%%%%%%%%%%%%%%%%%
%%%%%%%%%%%%%%%%%%%%%%%%%%%%%%%%%%%%%%%%%%%%%%%%%%%%%%%%%%

Let the class centroids and covariance matrices be defined as:
\begin{equation}
\mu_c = \mathbb{E}[\mathbf{v}|C=c], \quad 
\Sigma_c = \mathbb{E}[(\mathbf{v}-\mu_c)(\mathbf{v}-\mu_c)^T]
\end{equation}

The separability is quantified via the Fisher criterion:
\begin{equation}
J = \frac{\|\mu_1 - \mu_2\|^2}{\mathrm{Tr}(\Sigma_1 + \Sigma_2)}
\end{equation}

%%%%%%%%%%%%%%%%%%%%%%%%%%%%%%%%%%%%%%%%%%%%%%%%%%%%%%%%%%%%%
\subsection{Comparative Discussion: CNN vs. Hybrid STFT-ML}
%%%%%%%%%%%%%%%%%%%%%%%%%%%%%%%%%%%%%%%%%%%%%%%%%%%%%%%%%%%%%

%%%%%%%%%%%%%%% TABLE 4 %%%%%%%%%%%%%%%%%%%%%%%%%%%
\begin{table}[H]
\centering
\caption{Comparative Analysis and Feature Significance}
\label{tab:comparison}

\vspace{2pt}
\textbf{(a) Deep CNN vs.\ Proposed Hybrid STFT-ML}
\vspace{4pt}

\begin{tabular}{lll}
\toprule
\textbf{Metric} & \textbf{Deep CNN \cite{nogay2021}} & \textbf{Proposed} \\
\midrule
Input Dim. & $224 \times 224$ (Image) & $1 \times 10$ (Vector) \\
Architecture & AlexNet / ResNet & SVM-RBF \\
Parameters & $\sim$11.2M & $\sim$4K \\
Train Time & Hours (GPU) & 0.30 s (CPU) \\
Inf. Latency & $\sim$150 ms & 0.49 ms \\
Accuracy & 97.3\% & 94.4\% \\
Explainability & Low (Black-Box) & High (Physics) \\
Edge Deploy. & GPU Required & MCU Compatible \\
\bottomrule
\end{tabular}

\vspace{10pt}
\textbf{(b) Top-5 Feature Importance}
\vspace{4pt}

\begin{tabular}{clrl}
\toprule
\# & Feature & Score & Interpretation \\
\midrule
1 & RMS & 0.583 & Arc current magnitude \\
2 & HER & 0.306 & Harmonic ratio \\
3 & ZCR & 0.222 & Zero-crossing rate \\
4 & THD$_{arc}$ & 0.194 & Harmonic distortion \\
5 & $P_{100}^{n}$ & 0.167 & Normalized harmonic power \\
\bottomrule
\end{tabular}

\end{table}
%%%%%%%%%%%%%%%%%%%%%%%%%%%%%%%%%%%%%%%%%%%%%%%%%%%%%%%%%%

The computational efficiency advantage is expressed as:
\begin{equation}
\mathcal{O}_{CNN} \gg \mathcal{O}_{STFT-ML}
\end{equation}

%%%%%%%%%%%%%%%%%%%%%%%%%%%%%%%%%%%%%%%%%%%%%%%%%%%%%%%%%%%%%
\subsection{Physical Interpretability and Real-Time Tracking}
%%%%%%%%%%%%%%%%%%%%%%%%%%%%%%%%%%%%%%%%%%%%%%%%%%%%%%%%%%%%%%%

%%%%%%%%%%%%%%% FIGURE 5 %%%%%%%%%%%%%%%%%%%%%%%%%%%
\begin{figure}[H]
\centering
\includegraphics[width=1.1\columnwidth]{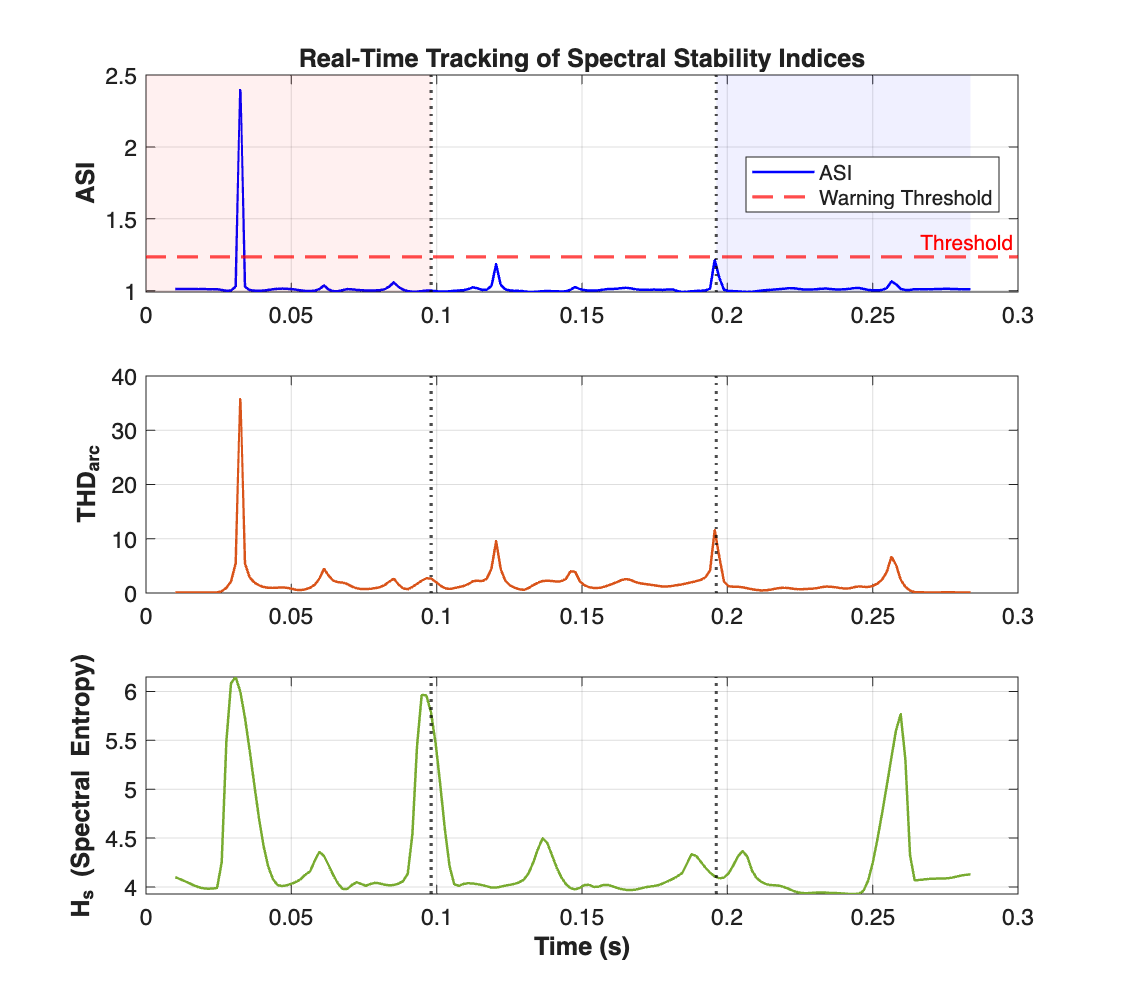}
\caption{Real-time tracking of spectral stability indices: ASI (top), THD$_{arc}$ (middle), and spectral entropy $H_s$ (bottom).}
\label{fig:asi_trend}
\end{figure}
%%%%%%%%%%%%%%%%%%%%%%%%%%%%%%%%%%%%%%%%%%%%%%%%%%%%%%%%%%

The evolution of spectral indices follows:

\begin{equation}
ASI(t) \sim e^{\lambda t}, \quad \lambda > 0
\end{equation}

indicating exponential instability growth prior to arc extinction.

Similarly:
\begin{equation}
H_s(t) \rightarrow H_{\max}
\end{equation}

suggesting maximum disorder.

%%%%%%%%%%%%%%%%%%%%%%%%%%%%%%%%%%%%%%%%%%%%%%%%%%%%%%%%%%%%%
%%%%%%%%%%%%%%% FIGURE 6 %%%%%%%%%%%%%%%%%%%%%%%%%%%
\begin{figure}[H]
\centering
\includegraphics[width=\columnwidth]{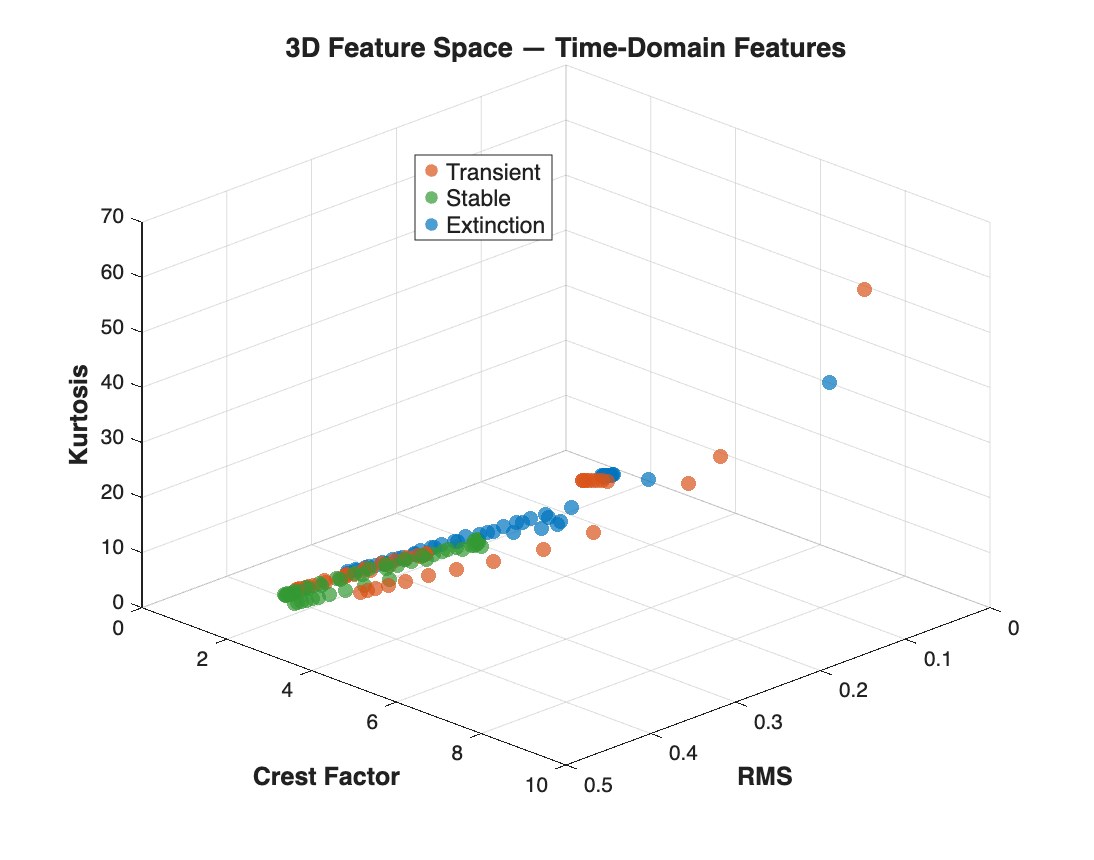}
\caption{Three-dimensional time-domain feature space (RMS, Crest Factor, Kurtosis).}
\label{fig:3d_time}
\end{figure}
%%%%%%%%%%%%%%%%%%%%%%%%%%%%%%%%%%%%%%%%%%%%%%%%%%%%%%%%%%

The combined feature space forms a manifold:
\begin{equation}
\mathcal{M} \subset \mathbb{R}^{10}
\end{equation}

on which each operational state corresponds to a distinct region, enabling interpretable classification.

%%%%%%%%%%%%%%%%%%%%%%%%%%%%%%%%%%%%%%%%%%%%%%%%%%%%%%%%%%%%%
\section{Limitations and Future Work}
%%%%%%%%%%%%%%%%%%%%%%%%%%%%%%%%%%%%%%%%%%%%%%%%%%%%%%%%%%%%%

Despite the promising performance of the proposed framework, several limitations should be acknowledged. First, the dataset used in this study is relatively limited in size, which may affect the generalization capability of the trained model. Although cross-validation and confidence interval analyses are conducted to mitigate this issue, further validation on larger and more diverse datasets is required.

Second, the evaluation results reveal a discrepancy between hold-out accuracy and cross-validation performance, indicating potential sensitivity of the model to variations in feature distribution. This is particularly evident in the transient regime, where the inherently non-stationary nature of the signal introduces additional complexity.

Third, the proposed feature representation, while physically interpretable, relies on predefined spectral bands and handcrafted descriptors. Although this enhances interpretability, it may limit adaptability to different welding conditions or signal characteristics.

Future work will focus on extending the dataset, improving robustness through adaptive feature learning, and validating the framework under real-time industrial operating conditions. In addition, integration with embedded systems and edge-based implementations will be explored to further assess practical deployment feasibility.

%%%%%%%%%%%%%%%%%%%%%%%%%%%%%%%%%%%%%%%%%%%%%%%
\section{Conclusion}
%%%%%%%%%%%%%%%%%%%%%%%%%%%%%%%%%%%%%%%%%%%%%%%

This study presented a physically-informed hybrid STFT-ML framework for arc stability monitoring in electric arc welding systems, where the measured current signal is interpreted as a stochastic representation of plasma dynamics. The proposed approach establishes a structured representation–inference pipeline that maps non-stationary signals into a compact and interpretable feature space derived from localized spectral energy distributions. Within this framework, the Arc Stability Index (ASI), spectral entropy ($H_s$), and harmonic distortion ($THD_{arc}$) are formulated as energy-based descriptors and integrated with time-domain features to capture both spectral redistribution and temporal variability.

The experimental results demonstrate strong classification performance, with a hold-out accuracy of 94.4\%. However, cross-validation results (85.6\% for Leave-One-Out and 87.5\% $\pm$ 9.4 for 10-fold) and the computed confidence interval (87.07\% with a 95\% CI of [81.65\%, 92.50\%]) provide a more realistic assessment of generalisation capability. The ROC and Precision–Recall analyses further confirm that the proposed feature space enables robust class separability, particularly for stable and extinction regimes, while transient states remain more challenging due to their inherently non-stationary behavior.

In contrast to high-dimensional deep learning approaches, the proposed framework achieves competitive performance with significantly lower computational complexity and inference latency, making it suitable for real-time deployment in resource-constrained environments. Although a modest reduction in peak accuracy is observed, this trade-off enables improved interpretability and a direct physical linkage between signal characteristics and diagnostic outcomes.

Overall, the main contribution of this work lies in the development of a physically interpretable and computationally efficient feature representation framework that bridges time–frequency analysis and machine learning-based classification. The results indicate that localized spectral energy redistribution can serve as a reliable indicator of arc stability, supporting practical condition monitoring and predictive maintenance strategies. Future work will focus on expanding the dataset, improving robustness under varying operating conditions, and validating the framework in real-time industrial environments.

\end{document}